\documentclass[twocolumn,preprintnumbers,aps,superscriptaddress,showpacs]{revtex4}

\usepackage{amsmath}
\usepackage{amssymb}

\input epsf
\input rotate

\include{dft}

\def\bea{\begin{eqnarray}}
\def\eea{\end{eqnarray}}
\def\ben{\begin{equation}}
\def\een{\end{equation}}
\def\benu{\begin{enumerate}}
\def\enu{\end{enumerate}}

\def\sss{\scriptscriptstyle\rm}

\def\1var{(\bx_1...\bx\N)}

\def\half{\frac{1}{2}}

\def\br{{\bf r}}

\def\b1{{\bf 1}}
\def\bx{{x}}

\def\T{^{\sss TDDFT}}
\def\K{^{\sss KS}}

\def\s{_{\sss S}}
\def\xc{_{\sss XC}}

\def\Hxc{_{\sss HXC}}

\def\N{_{\sss N}}
\def\H{_{\sss H}}

\def\ext{_{\rm ext}}

\def\ee{_{\rm ee}}

\def\sph_int{ {\int d^3 r}}

\def\infintd3r{ \int_{-\infty}^\infty d^3r\,}
\def\intd3r{ \int d^3r\,}

\def\laplace1d{\frac{d^2}{dx^2}}
\def\plaplace1d{\frac{d^2}{d{x'}^2}}

\def\padr2{\frac{\partial^2}{\partial r^2}}

\def\N{{\cal N}}

\begin{document}
\title{Investigating interaction-induced chaos using time-dependent density functional theory}
\date{\today}
\author{Adam Wasserman}
\affiliation{Department of Chemistry and Chemical Biology, Harvard
University, 12 Oxford St. Cambridge, MA 02138, USA.}
\author{Neepa T. Maitra}
\affiliation{Department of Physics and Astronomy, Hunter College
and City University of New York, 695 Park Av. New York, NY 10021,
USA.}
\author{Eric J. Heller}
\affiliation{Department of Chemistry and Chemical Biology, Harvard
University, 12 Oxford St. Cambridge, MA 02138, USA.}
\affiliation{Department of Physics, Harvard University, 17 Oxford
St. Cambridge, MA 02138, USA.}

\begin{abstract}
Systems whose underlying classical dynamics are chaotic exhibit
signatures of the chaos in their quantum mechanics. We investigate
the possibility of using time-dependent density functional theory
(TDDFT) to study the case when chaos is induced by
electron-interaction alone.  Nearest-neighbour level-spacing
statistics are in principle exactly and directly accessible from TDDFT.
 We discuss how
the TDDFT linear response procedure can reveal the {\it mechanism}
of chaos induced by electron-interaction alone. A simple model of
a two-electron quantum dot highlights the necessity to go beyond
the adiabatic approximation in TDDFT.
\end{abstract}

\maketitle

\section{Introduction}
The study of the quantum mechanics of systems whose underlying
classical dynamics are chaotic, has revealed many intriguing
features. It is now well-established that underlying classical chaos
dramatically manifests itself in certain quantum signatures: spectral
fluctuations, localization properties of wavefunctions, inverse participation ratios, and extreme
sensitivity to tiny variations in control
parameters~\cite{Haake}. Perhaps the most widely studied of these is
the spectral statistics: In particular, the nearest-neighbour spacing
distributions (NNS) for integrable systems generically display Poissonian
statistics (level clustering), while displaying Wigner-Dyson
statistics (level repulsion) for chaotic systems. Such correspondence
between classical chaos and NNS statistics was conjectured by Bohigas, Giannoni, and Schmit 
\cite{BGS84}, and while counterexamples exist \cite{WVFS90,ZDD95}, the association is so 
general that Wigner-Dyson statistics are often viewed as signatures of underlying classical chaos.
 This property has
been used to study the transition from integrability to chaos as a
parameter in the confining potential of the system is
varied~\cite{BohigasLesHouches}.
Many of these studies consider one particle, or, in quantum dots,
non-interacting electrons; the chaos arises in such cases due to
the shape of the dot confining potential.  What is less
well-understood is what happens when electron-interaction
is turned on. Suppose that interacting electrons are placed in a potential where the single-particle NNS
is Poissonian; is the chaos induced by the Coulomb interaction enough
to transform the NNS statistics to a Wigner-Dyson distribution?
To what extent does the level repulsion kick in? Is the picture
qualitatively different with few electrons compared to many
electrons?

That electron-interaction alone can induce chaos is certainly
evident from the very early days of quantum mechanics, impeding
Bohr in 1913 from successfully quantizing the Helium atom
\cite{B13} (task only completed semiclassically in 1991
\cite{ERTW91}; for a beautiful review of the theory of
two-electron atoms, see ref.\cite{TRR00}). Putting aside concerns
regarding the validity of random-matrix theory when 2-body
interactions are present \cite{GMW98}, we note that the first
application of random-matrix theory outside of nuclear physics was
to a series of complex atoms~\cite{RP60, T61}, where the
spin-orbit interaction provided the crucial ingredient in yielding
Wigner-Dyson statistics. Statistics of a different series of
complex atoms based on experimental data~\cite{CG83}, as well as
theoretical models~\cite{FGGK94} support the finding that highly
excited states of complex atoms tend to display Wigner-Dyson
statistics. There have been several studies in
molecules~(eg.~\cite{ZKCP88}) where the coupling of electronic
excitations with nuclear vibrational and rotational excitations
provide the complexity. Generally, experimental data, with many
levels of the same symmetry, is needed. The chaos in these
examples is understood to arise not from Coulomb
electron-interaction alone but rather from its coupling to other
degrees of freedom.

For simple atoms, there have been very few calculations of the
level-spacing statistics, perhaps because of the challenges involved
in gathering enough levels of doubly-excited resonances, either from
theory or experiment, which appear to be crucial for this effect (see
Ref.~\cite{PGDD00} for a calculation of level-repulsion in
helium). However deviations from Poissonian statistics have been
clearly identified in non-hydrogenic Rydberg atoms in either magnetic
or electric fields~\cite{JGD98,MM01,HS00,KKVH99}. The hydrogen atom at
the corresponding parameter regime is integrable, but scattering off
the ionic core in a non-hydrogenic Rydberg atom creates fundamentally
different dynamics, with chaotic trajectories depending on the value
of the quantum defect. In models of this effect, NNS have been shown
to display level-repulsion, following a distribution intermediate
between Poissonian and Wigner-Dyson~\cite{JGD98}.

There are also many-electron solid-state examples, where
electron-interaction has shown to lead to the transition between
Poissonian and Wigner-Dyson statistics~\cite{JS97,AGKL97,BA96,PM98},
and several of these works identify the driving parameter for this
transition.
Difficulties with interpreting experimental data make the idea of
a theoretical calculation of the level statistics attractive. At
the same time, because the solution of the interacting
many-electron problem grows exponentially with the number of
electrons, typically a model is used for the interaction, eg. a
two-body random interaction model was used in Ref.~\cite{JS97}.  A
recent calculation of just two electrons in a quantum
dot~\cite{FST01}, explicitly demonstrates that Coulomb
electron-interaction alone can transform the Poissonian statistics
of the non-interacting system into Wigner-Dyson.  Generally in
few-electron quantum dot studies of transport \cite{FLP03}, random
constant-interaction models~\cite{A00} are often used, however
Ref.~\cite{FST01} suggests this may not be a good description of
the real interacting electronic system, as constant-interaction
models retain the Poissonian statistics of the integrable
non-interacting system.

In order to better understand the mechanisms of interaction-induced
chaos, it would be desirable to use a method that captures electron
correlation reliably and efficiently (given the thousands of excited
states needed in the calculation), scales well with the number of
electrons, and from whose procedure one can glean aspects of the
mechanism that brings about the chaos.

Time-dependent density functional theory (TDDFT) is the leading
candidate for such a method.  This has become the method of choice for
the calculation of excitations and response properties of interacting
electronic systems, because of its scalability: typically the accuracy
is comparable to sophisticated wavefunction methods such as Complete Active
Space Self-Consistent Field (CASSCF),
while implementations follow a far cheaper scaling with system size,
comparable to time-dependent Hartree-Fock (TDHF). The theory in principle yields exact excitations, but
approximations for exchange-correlation effects are needed in
practise. Typically, excitation energies are given to within a few
tenths of an eV, although there are notorious exceptions~\cite{BWG05,MUNR06}. There has been a
tremendous drive in recent years to develop and improve the
currently-available functionals.

In this paper we will explore the possibilities of using TDDFT for
investigating quantum chaos induced by electron-interaction
alone. This is a new area for TDDFT. The idea is to use TDDFT to study
the transition from clustering to repulsion statistics in a given
system as the electron interaction is turned on. For the case of chaotic quantum dots, ground-state DFT has been used within the local spin density approximation
to study the statistics of ground-state spin and spacing between conductance peaks in the Coulomb blockade regime (addtion spectra) \cite{JUYB04,UJYB05}, but to our knowledge, the excited-state statistics of isolated dots with fixed number of electrons have not been studied via density functional methods.
 We will show in
Section~\ref{sec: TDDFT} how, by monitoring the evolution of the spectral statistics at different stages of the calculation, the TDDFT linear response framework can
also shed light on the {\em mechanism} of interaction-induced chaos.

A most essential question is whether the present-day functionals are
good enough to perform these tasks. In Sec.~\ref{sec:Model} we will
give an example of a case where they are not. This example highlights
an important challenge that must be overcome for TDDFT to be used in
these studies.

Aside from fundamental interest, it is important to characterize
chaotic versus integrable dynamics for applications of technological
interest, such as transport across quantum dots, highly excited atoms and molecules in external fields, quantum control and
manipulation in external fields, or the engineering of quantum computer hardware that maximizes the fidelity of quantum computations \cite{LS06}.

\section{Level Statistics from Time-Dependent Density Functional Theory}
\label{sec: TDDFT}

TDDFT \cite{RG84} has
become the method of choice to calculate a variety of response
properties of molecules, clusters, and solids, in the presence or
absence of external time-dependent fields \cite{MUNR06}.  The
Runge-Gross theorem provides a rigorously exact foundation for the
theory: this states that given the initial state of the interacting
electronic system, {\it all} observables of the system can be
extracted in principle from just the time-evolving density.

Most applications in
chemistry and solid-state physics currently fall in the linear
response regime, where TDDFT yields predictions for the optical
spectra, i.e. the frequency and intensity of electronic
excitations in response to  electric fields
\cite{PGG96,C96}.

We now explain in some detail the standard computational procedure.
Consider a time-independent $N$-electron Hamiltonian (atomic units will be used throughout)
\ben
\hat{H} = \hat{T} +\hat{V}\ee + \int d\br \hat{n}(\br)v\ext(\br)~~,
\label{eq:Hamiltonian}
\een
whose energy spectrum $\{E_k\}$ we want to calculate. In Eq.(\ref{eq:Hamiltonian})
$\hat{T}=-\half\sum_i\nabla_i^2$ stands for the $N$-electron kinetic energy operator,
$\hat{V}\ee=\half\sum_{i,j\neq i}|\br_i-\br_j|^{-1}$ for the electron-electron repulsion, $v\ext(\br)$ for the external potential due
to the nuclei, or applied static fields, and $\hat{n}(\br)=\sum_{i=1}^N\delta(\br-\hat{\br}_i)$ is
the density operator. The {\em first step} of a
linear-response TDDFT calculation involves the self-consistent
solution of the {\em ground-state} Kohn-Sham (KS) equations \cite{KS65}:
\ben
\left(-\frac{1}{2}\nabla^2+v_s[n](\br)\right)\phi_i(\br)=\epsilon_i\phi_i(\br)\label{eq:KS_eqn}
\een
Here $v_s[n](\br)$ is the KS potential defined such that $N$ non-interacting electrons in $v_s[n](\br)$ have the same ground-state density $n(\br)=\langle\Psi_0|\hat{n}(\br)|\Psi_0\rangle$ as the original interacting system of
ground state $|\Psi_0\rangle$. In the KS scheme, the density is obtained from the $N$ occupied KS orbitals as $n(\br)=\sum_{i~occ}|\phi_i(\br)|^2$. The KS potential is written as the sum of three pieces:
\ben
v\s[n](\br) = v\ext(\br) + v\H[n](\br) + v\xc[n](\br)~~,
\label{eq:vs}
\een
where $v\H[n](\br) = \int d^3r' n(\br')/\vert \br - \br'\vert$ is the Hartree potential, and $v\xc[n](\br)$ the
exchange-correlation potential. This is the functional derivative (with
respect to the density) of the exchange-correlation energy functional
$E\xc[n]$, evaluated at the ground-state density:
$v\xc[n](\br)=\left.\delta E\xc[n]/\delta n(\br)\right|_{n}$ . The
functional $E\xc[n]$ is the only quantity that needs to be
approximated in order to get the ground-state density and energy; it
is fortunately amenable to local
approximations~\cite{PK05} in many situations. Knowledge of $v\xc(\br)$ is sufficient to calculate the ground-state energy $E_0$, 
but the excited-state energies $\{E_k\}$, $k>0$, are not accessible from this first step.
The occupied orbital energies $\epsilon_i$, and KS orbitals $\phi_i(\br)$, along with the unoccupied orbital energies and orbitals,
 are used as basic
 ingredients for the {\em second step} of the calculation. The aim of this second step is to correct
the unphysical KS excitations towards the correct ones.
Here one solves for eigenvalues and eigenvectors of the matrix:
\ben\tilde{\Omega}_{qq'}=\delta_{qq'}\omega_q^2+4\sqrt{\omega_q\omega_{q'}}
\left[q\left|f\Hxc(\omega)\right|q'\right]~~.\label{e:casida}\een
The square-root of the eigenvalues correspond to the excitation frequencies $E_k-E_0$.
The double index $q=(i,a)$ represents a single excitation from
occupied KS orbital $\phi_i(\br)$ to unoccupied KS orbital
$\phi_a(\br)$, $\omega_q$ is the difference between occupied and
unoccupied KS orbital energies, $\omega_q=\epsilon_a-\epsilon_i$,
and
\begin{eqnarray}\nonumber
\left[q\left|f\Hxc(\omega)\right|q'\right]=&&\int d\br d\br'
\phi_i^*(\br)\phi_a(\br)\\&&\times
f\Hxc(\br,\br';\omega)\phi_{i'}(\br')\phi_{a'}^*(\br')~~.\label{e:expectation_kernel}\end{eqnarray}
The Hartree-exchange-correlation kernel,
$f\Hxc[n](\br,\br';\omega)$, is the central quantity of
linear-response TDDFT. In the time-domain, $f\Hxc[n](\br
t,\br't')$ is the functional derivative of the {\em
time-dependent} Hartree-plus-exchange-correlation potential
$v\H[n_t](\br,t) + v\xc[n_t](\br,t)$ with respect to the
time-dependent density $n_t(\br't')$, evaluated at the
ground-state density $n(\br')$(see Sec.4.3.2 of \cite{MG03} for
more details). Eqs.~(\ref{e:casida}) and
~(\ref{e:expectation_kernel}) are obtained from the linear
response limit of the full TDDFT equations, in which $N$ electrons
evolve in the time-dependent KS potential \ben v\s[n_t](\br,t) =
v\ext(\br,t) + v\H[n_t](\br,t) + v\xc[n_t](\br,t)~~. \een Here,
$n_t=n(\br,t)$ is the time-dependent density of the $N$ KS
electrons, which is the same as that of the interacting
system~\cite{RG84}.

To summarize, there are two stages:
(1$^{\rm st}$) The ground-state Kohn-Sham (KS) equations
are solved to self-consistency; this requires an
approximation to the ground-state exchange-correlation energy
functional. Even if the (unknown) ``exact" functional were used
here, the excitations of the ground-state KS potential (i.e differences between occupied and unoccupied KS orbital
energies) could only be regarded as zeroth-order approximations to the
true excitations of the system.
(2$^{\rm nd}$) The KS frequencies
are corrected via $f\Hxc$ to become the true
excitations of the many-body system \cite{PGG96}.

With rare exceptions, an adiabatic approximation (ATDDFT) is employed
in this second step: the exchange-correlation potential at time $t$ is
approximated by that of a ground-state of the instantaneous density
$n(\br t)$.  This means that the Hartree-exchange-correlation kernel
has no frequency-dependence in ATDDFT. An important consequence for
the purposes of this paper is that states of double (or higher
multiple) excitations cannot be captured within
ATDDFT~\cite{TH00,MZCB04}; such excitations require a strongly
frequency-dependent kernel.  In the simplest (and common) adiabatic local-density approximation (ALDA), the
approximation is local in space as well as time.

Notwithstanding the resounding success of TDDFT within existing
approximations, different instances where the approximations fail
have been identified. Charge-transfer excitations at long-range,
conical intersections, states of multiple-excitation character,
polarizabilities of long-chain polymers, Rydberg excitations, lie
among the challenges in the linear response regime; non-sequential
photoionization and quantum control applications are important
challenges for approximations in the strong-field regime. Given
the success of TDDFT for the vast majority of problems, and the
fact that it scales in a reasonable way
with system size while incorporating electron-correlation effects,
there is a tremendous drive in the recent literature to understand
and improve TDDFT approximations.

\vspace{0.5cm}

\textbf{Elucidating the mechanism of interaction-induced chaos: type-KS vs. type-$f\Hxc$ chaos}

Consider now a system of $N$ interacting electrons with
Wigner-Dyson NNS statistics, and focus on the case where the
external potential $v\ext(\br)$ is such that the classical
dynamics of a single electron in $v\ext(\br)$ is integrable. The
statistics of $N$ non-interacting electrons moving in $v\ext(\br)$
follows that of the one-particle system, i.e. Poissonian in this
integrable case. Suppose also that the {\em exact} functionals
$E\xc[n]$ and $f\xc[n](\br,\br';\omega)$ are known. The exact
TDDFT procedure must transform the Poissonian non-interacting
statistics to the correct interacting ones. In what follows, when
the interacting system displays Wigner-Dyson statistics, we refer
to this as the P$\to$WD transition. Moreover, in studying {\em
how} it does so, the mechanism for interaction-induced chaos can
be better understood.

First, the ground-state KS potential $v_s(\br)$ is found, and we
ask: Is the classical dynamics of a single electron in $v_s(\br)$
already chaotic? In this scenario, at least part of the
interaction-induced chaos appears in DFT as a ``chaotic kink" in
the Hartree and/or exchange-correlation pieces of the potential.
The ``kink" is that piece of the KS potential due to which the
single-particle classical dynamics is chaotic. Excitations from
the bare KS potential will generally show in this case some degree
of level repulsion, perhaps enough to agree with the experimental
NNS distribution. We refer to this as type-KS chaos.

(We note here that it is preferable to perform the level
statistics in such a way that $N$ non-interacting electrons in a
given potential follow the same type of distribution as a single
electron in that potential: this can be achieved if the statistics
are performed on energy levels of a fixed symmetry class. That is,
if there are constants of the motion in addition to the total
energy, one fixes the value of each of those constants. In the
case of $N$ non-interacting electrons, each orbital eigenvalue is
a conserved quantity; therefore one fixes all except one, since
the sum of the orbital energies gives the total energy. This
amounts to considering only single excitations of the system;
these can be out of any of the occupied orbitals, depending on
which were chosen fixed but the resulting statistics will be
independent of the choice).

In the second step of TDDFT, the bare KS excitations are corrected to
the true ones using $f\Hxc$: What is its effect on the statistics?
Certainly if the KS system turned out to be integrable, then the
entire job of transforming the statistics is done by the exact
$f\Hxc$. We refer to this as type-$f\Hxc$ chaos.

For $N$ electrons in one dimension, type-KS chaos is impossible. The
dynamics of a single electron in one dimension is obviously always
integrable, since there is one conserved quantity (the energy of the
electron) and one degree of freedom. Using the terminology
above, the Kohn-Sham potential will never acquire a ``chaotic kink" in
one dimension. In more than one dimension the modulation of the external
potential that is provided by the Hartree and xc terms
(Eq.~(\ref{eq:vs})) may induce chaotic motion on the one-electron
dynamics. Nevertheless, as we will discuss in Section
\ref{sec:Discussion}, type-$f\Hxc$ chaos is likely to be more common  also in more than one
dimension. Generally, most of the job required for the P$\to$ WD
transition has to be done by $f\Hxc$.

Of course in practise, approximate functionals must be used. The
question then arises: what properties are needed in approximate
kernels to achieve the correct statistics?  Are the present-day
approximations good enough to capture the P$\to$WD transition as the
interaction is turned on? In particular, the adiabatic approximation
for the kernel: this mixes KS single-excitations but cannot fold in
multiple excitations~\cite{TH00,MZCB04}. This immediately raises a red
flag as a chunk of the true excitations are excluded from
consideration: how significant is this chunk? Even if the adiabatic
kernel creates level repulsion by mixing singles only, if the
double-excitations compose a significant proportion of the spectra in a
certain range, then the ATDDFT spectra would not be
representative of the true system. The answers to these questions are
likely system-dependent. We illustrate in the next Section most of the
concepts discussed so far using a model of a one-dimensional quantum
dot.


\section{A Model 1-d Quantum Dot}
\label{sec:Model}
Consider
the problem of two electrons interacting via a soft-Coulomb
potential $v_{ee}(x_1,x_2)=(1+(x_1-x_2)^2)^{-1/2}$ in a
one-dimensional box of length $L$. A simple analysis shows that
the matrix elements of $v_{ee}$ scale approximately as $1/L$ (basis of
non-interacting electrons), whereas the kinetic-energy matrix
elements scale as $1/L^2$. The interaction in this sense becomes
more important as the length of the box is increased. The
non-interacting system is of course integrable, having two
constants of the motion (the energies of the individual particles)
for two degrees of freedom. The weakly-interacting case (small
$L$) is then almost-integrable, and the strongly-interacting case
(large-$L$) non-integrable. In fact, ref.\cite{FST01} shows clear
signatures of interaction-induced chaos in a similar system. We
start, however, in the weakly-interacting limit, realized for
$L=1$ a.u.

\begin{figure}
\begin{center}
\epsfxsize=90mm \epsfbox{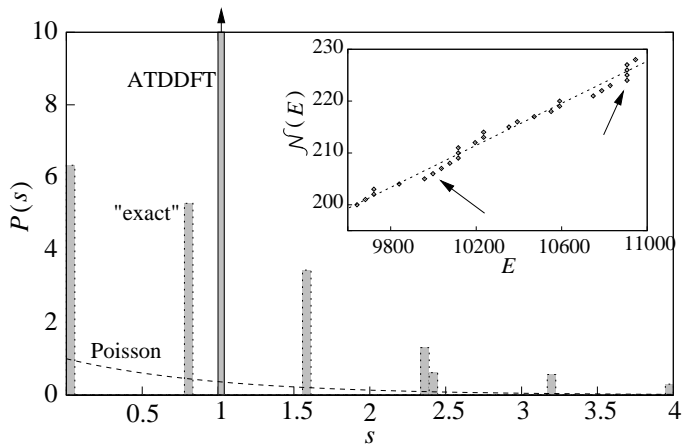} \caption{Light-shaded
bars: NNS histogram for the exact spectrum of 2 interacting
electrons in a 1-d box of length $L=1$ a.u. Dark-shaded bar:
ATDDFT spectrum within exact-exchange. Top arrow indicates that the bar
keeps going up. Inset: small segment of staircase function
$\N(E)$. The dotted line shows the smooth $\overline\N(E)$. Arrows
point to the only two single excitations in this
region.}\label{f:L1}
\end{center}
\end{figure}

\begin{figure}
\begin{center}
\epsfxsize=90mm \epsfbox{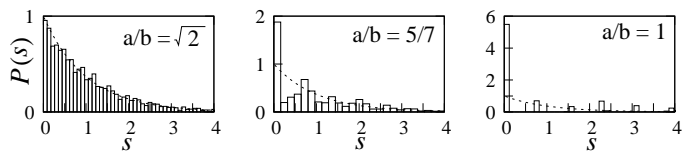} \caption{The NNS
distribution for a particle in a 2-d rectangular billiard of sides
$a^2$ and $b^2$ depends on the $a/b$ ratio as indicated here.
The dashed line corresponds to the Poisson distribution. (Shown first in ref.\cite{BT77}).}\label{f:non_int}
\end{center}
\end{figure}

The light-shaded bars of Fig.(\ref{f:L1}) correspond to the
nearest-neighbor-spacing (NNS) distribution $P(s)$ obtained via
the following three steps: (1) Get the first $N_{\rm max}$ levels
of the energy spectrum $\{E_i\}$ by exact diagonalization of the
2-electron Hamiltonian, and discard the lowest $N_{\rm min}$. We
typically used $N_{\rm max}=2000$, $N_{\rm min}=200$, and only
states of even symmetry; (2) Unfold the staircase function
$\N(E)$, representing the number of states having energy less than
$E$, by applying the map $x_i=\overline{\N}(E_i)$, where
$\overline{\N}$ is the smooth part of $\N$, found via a 4$^{\rm
th}$-order polynomial fit to $\N(E)$; (3) Set $s_i=(x_{i+1}-x_i)$,
and count $P(s)$, the number of occurrences for each
$s_i/\overline{s}$, where $\overline{s}=\sum_i s_i/(N_{\rm
max}-N_{\rm min})$. $P(s)$ then contains information about
inherent fluctuations of the level spacings. As discussed in the introduction, these are generally
Poissonian for integrable systems, and Wigner-Dyson-like for
chaotic systems \cite{BFFM81}, but may also be non-generic for some highly symmetric systems like the one we are considering here.
The exact histogram consists of a series of bars separated by
gaps.
The fluctuations around $\overline\N(E)$ (dotted line on the inset
of Fig.(\ref{f:L1})) are evidently not random; they exhibit
distinctive patterns, each of which gives rise to a different bar
in the histogram. In the non-interacting limit, the problem of two
particles with the same mass in a 1-dimensional box is equivalent
to that of a single particle in a 2-dimensional square box, and
the patterns just mentioned arise from the square symmetry of the
problem. These remain in the weakly-interacting regime. The
histogram would be more Poissonian if the two particles had
different masses, and even more so if the ratio of the two masses
were an irrational number, as shown in Fig.(\ref{f:non_int}) (and
ref.\cite{BT77}).

What does TDDFT predict for the histogram within the adiabatic
approximation?: a single bar centered at $s=1$ (dark-shaded bar in
Fig.\ref{f:L1}). We now explain why.


\begin{figure}
\begin{center}
\epsfxsize=70mm \epsfbox{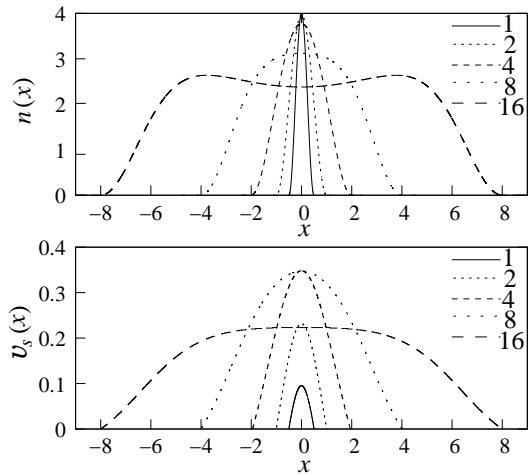} \caption{Ground state
density $n(x)$ and Kohn-Sham potential $v_s(x)$ for 2 interacting
electrons in a 1d-box, as its length is increased from $L=1$ to
$L=16$. The walls of the boxes at $\pm L/2$ are not
shown.}\label{f:dens_pot}
\end{center}
\end{figure}

As discussed in Section \ref{sec: TDDFT}, the Kohn-Sham scheme
transforms the original problem of 2 interacting electrons into that
of a single electron moving in the potential
$v_s(x)=v\ext(x)+v\Hxc(x)$, where $v\ext(x)$ is the external potential
(box of unit length, in this case).  Being a one-dimensional
potential, it cannot exemplify type-KS chaos. The
Hartree-exchange-correlation potential, $v\Hxc(x)$, was found here
within the exact-exchange approximation \cite{KLI92}, as implemented in the octopus code~\cite{MCBR03}. $v\Hxc(x)$ is in this
case simply a small bump at the bottom of the box (see solid line in
lower panel of Fig.(\ref{f:dens_pot})), having practically no
influence on the high-energy spectrum of the non-interacting problem,
except for a small shift in all the high-lying energies, irrelevant
when the analysis of the {\em differences} between neighboring levels
is made. The Kohn-Sham equations (Eq.\ref{eq:KS_eqn}) yield a set of
orbital energies $\{\epsilon_i\}$, but only the subset of {\em
occupied} orbitals is used to get the ground-state energy of the
interacting system, as: $E_0=\sum_{i~occ}\epsilon_i+E\Hxc[n]-\int dx
n(x)v\Hxc(x)$. The excited-state energies $E_n\T$ are obtained within
linear-response TDDFT as:
\ben
E_n\T=E_0+\omega_n\T~~,\label{e:total_energy}\een where the
squares of $\omega_n\T$, $\Omega_n\equiv\left(\omega_n\T\right)^2$
are the eigenvalues of the matrix of Eq.\ref{e:casida}.

The $f\Hxc$ kernel we employed is again that of exact-exchange, local in
time (proportional to $\delta(t-t')$). Its Fourier transform with
respect to $t-t'$ is therefore frequency-independent. As a
consequence, from Eq.(\ref{e:casida}) and
Refs.~\cite{TH00,MZCB04}, ATDDFT only yields corrections to KS
single-excitations. Since the frequencies $\omega\T_n$ of
Eq.(\ref{e:total_energy}) are all mixtures of single excitations,
the ATDDFT spectrum at this level contains no double excitations
at all (or multiple excitations for $N>2$). This is a serious
problem even at the low energies that are of interest for most
quantum-chemical purposes, but in the present setting, the missing
of double excitations is simply catastrophic. Single excitations
are not more than a negligible fraction of the high-energy set of
levels. The arrows of the inset in Fig.(\ref{f:L1}), for example,
point to the only two single excitations found in the
lowest-energy segment taken into account for our statistical
analysis.

We are only one step away from concluding that in the energy range
entering the statistical analysis, the ATDDFT staircase function
is identical to the bare KS staircase function: \ben \N^{\sss
ATDDFT}(E)=\N\K(E)~~. \label{e:staircase_equality}\een To see
this, we just point out that the expectation value of
Eq.(\ref{e:expectation_kernel}) becomes vanishingly small in the
adiabatic approximation for high values of $\omega_q$. It can be
seen in Table {\ref{t:table1}} that for unoccupied orbital indices
$a\gtrsim 10$, adiabatic single-excitations barely differ from the
KS ones on the scale of energy fluctuations.

Furthermore, in the weakly-interacting limit, $\N\K(E)$ coincides
with the staircase function for a {\em single electron} in the
presence of $v\ext(x)$, since, as argued before, the effect of
$v\Hxc(x)$ is negligible. $\N\K(E)$ has no fluctuations at all in this limit, being
precisely proportional to $\sqrt{2E}$, and the unfolding process
translates this into a histogram showing only $s=1$ spacings.

It may be argued that double excitations can simply be {\em added}
to the ATDDFT spectrum by including the sums of
$\epsilon_a+\epsilon_b$ with $a$ and $b$ running over all the
unoccupied orbitals. This seems entirely sensible, and works of
course in the non-interacting limit, but the adiabatic approximation yields no
corrections to these states, and Eq.(\ref{e:staircase_equality})
still holds true.

\begin{figure}
\begin{center}
\epsfxsize=90mm \epsfbox{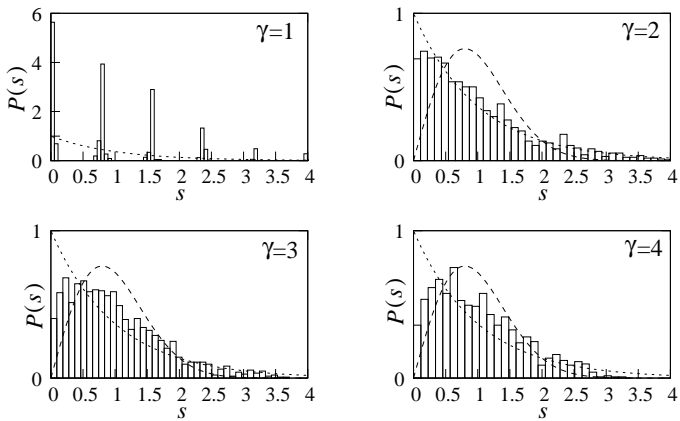}
\caption{NNS histograms for 2 interacting electrons in a 1d-box of
length $L=10^\gamma$. $P(s)$ approahes a Wigner-Dyson distribution
(dashed) for very wide boxes, a signature of chaos in the
underlying classical dynamics. The Poisson distribution is
indicated by dotted lines.}\label{f:L_dependence}
\end{center}
\end{figure}

We now discuss what occurs as we move towards the
strongly-interacting regime. Fig.(\ref{f:L_dependence}) shows the
NNS histograms as the length of the box is increased from $L=10$
to $L=10^4$.
Strong level repulsion manifests itself with a marked decrease of
small spacings. The histogram tends to a Wigner-Dyson distribution
$P(s)=\frac{\pi}{2}se^{-(\pi/4)s^2}$ for very wide boxes. This is a quantum signature of the underlying interaction-induced chaos, as implied by the classical dynamics results obtained by Fendrik et.al.\cite{FST01}, showing that apart from small regular regions in phase space, the dynamics of the system can become strongly chaotic due to electron-electron interactions. But it follows from
Eq.(\ref{e:staircase_equality}) that ATDDFT yields no level
repulsion. Practically all levels entering the statistical
analysis are double excitations. Even if we decided to
artificially add these by summing unoccupied KS orbital energies,
as discussed before, only a mixture of uncorrelated levels would
result, i.e. a Poisson distribution $P(s)=e^{-s}$ with no level
repulsion whatsoever (dotted lines in Fig.(\ref{f:L_dependence})).
This important signature of interaction-induced chaos is simply
not captured by ATDDFT.

As
the length of the box increases, localization of the electrons in
opposite extremes of the box starts taking place, and convergence
of the KS equations becomes problematic (we used the Octopus code \cite{MCBR03} for
the DFT calculations). Fig.(\ref{f:dens_pot})
shows the evolution of the KS potential and ground-state density
as $L$ is increased from 1 to 16.
The problems here are no different than
those encountered when trying to describe the formation of a
Wigner crystal \cite{SZRS78} or other phenomena in the
strongly-correlated regime. There
is already a hint of localization when $L=16$ (long-dashed line,
upper panel of Fig.\ref{f:dens_pot}). The KS potential adopts the
shape of a double well, with $v\Hxc(x)$ as the barrier in the
middle, growing stronger with respect to level spacings as the
length of the box increases.





\begin{table}
\caption{Comparison of low-frequency excitations (in Hartrees)
between bare Kohn-Sham energy differenes $\epsilon_a-\epsilon_i$
and TDDFT frequencies obtained via Eq.(\ref{e:casida}) with the exact-exchange approximation
to $v\xc$ and an adiabatic aproximation to $f\xc$.}
\begin{center}
\begin{tabular}{|c|c|c|}
\hline
$i\to a$&KS&TDDFT\\
\hline
1 $\to$ 2 & 0.0939 & 0.2771\\
1 $\to$ 3 & 0.3362 & 0.5132\\
1 $\to$ 4 & 0.6736 & 0.8014\\
1 $\to$ 5 & 1.1091 & 1.1963\\
1 $\to$ 6 & 1.6425 & 1.7015\\
1 $\to$ 10 & 1.7496 & 1.7628\\
1 $\to$ 20 & 1.9319 & 1.9319\\
\hline
\end{tabular}
\end{center}
\label{t:table1}
\end{table}


\section{Discussion and Outlook}
\label{sec:Discussion}

Our example model system was an unfortunate one for the usual
adiabatic approximations of TDDFT. The missing double excitations
haunt the problem even in the integrable limit, making up the
significant fraction of excitations in the true space and being
crucial to bringing about the chaos. There is intensive
development to go beyond the adiabatic approximation in TDDFT (eg.
Refs~\cite{VK96,T05,YB05,MZCB04}), and the progress has been
successful in many cases. It will be interesting to adapt the
kernel of Ref.~\cite{MZCB04,CZMB04} to the case in the current
paper.

We can now justify our claim of Section \ref{sec: TDDFT} that
type-KS chaos is rare. Even in cases where $v\Hxc(\br)$ develops
a ``chaotic kink" (see second part of Sec.\ref{sec: TDDFT}), level
repulsion in the bare Kohn-Sham system will only be observed when the relevant excitations that
enter the statistical analysis are of a single-particle nature. In
such cases, ATDDFT should perform much better than in the
model of Section \ref{sec:Model}. The spectral properties of
nonhydrogenic atoms in weak external electric or magnetic fields
seem to be suitable examples. The joint effect of an ionic core of
inner electrons (describable by a quantum defect for a series of
single-excitations) and the external field, has been shown to lead
to chaos in some parameter regimes \cite{JGD98}. The fact that
accurate quantum defects can be obtained from bare Kohn-Sham
potentials, and that the external field appears explicitly in the
KS equations, suggests that ATDDFT has a good chance to
succeed in the description of this phenomenon. It will be
interesting to compare the ATDDFT spectrum with that of
the scatterer-perturber models in the literature,
Refs.~\cite{JGD98,HS00,KKVH99,MM01}. TDDFT treats all
electrons quantum mechanically, and provides an {\it ab initio}
method against which to compare the model potentials.

It should also be noted that the level-spacing statistics physically
affect very high-temperature properties of the system.  Other measures
of chaos such as localization properties of wavefunctions are more
relevant at usual lower temperatures.  Before TDDFT can be used for
these properties, one must first solve the TDDFT ``observable
problem'' i.e. quantities that are directly given by the time-evolving
density (or in perturbation theory, the density response), such as the
excitation energies, oscillator strengths, or dipole moments, are
simply extracted from the Kohn-Sham system. However, for properties
that rely on other aspects of the interacting wavefunction, such as
its localization in phase-space, the information is not readily
accessible from the Kohn-Sham system. Such cases demand the
construction of appropriate functionals of the time-dependent density
for the relevant wavefunction-dependent observable, often a challenging task (see, for example, Ref.~\cite{WB06}).

In summary, we have discussed prospects for using TDDFT to study
how electron-interaction induces chaos in an otherwise integrable
system. The scalability of TDDFT and its success in describing
electron correlation in an ever-widening range of problems, make
it attractive for this. Moreover, {\it in principle}, TDDFT yields
exact level statistics for the interacting electronic system.  We
have discussed how the majority of currently-available approximate
functionals however cannot capture the correct statistics in many
cases of interest in quantum chaos, because of their inability to
describe excited states of multiple-excitation character.
Therefore challenges lie in developing approximate and
easily-implementable functionals for this purpose.

{\it Acknowledgement} We thank Jamal Sakhr for useful discussions.
NTM is grateful for financial support from the National Science
Foundation CAREER program, CHE-0547913.

\end{document}